\documentclass[12pt]{article}
\usepackage{amssymb,amsmath,epsfig}
\allowdisplaybreaks

\begin{document}
\title{\bf Noether Symmetry Approach in Energy-Momentum Squared Gravity}
\author{M. Sharif \thanks {msharif.math@pu.edu.pk} and M. Zeeshan Gul
\thanks{mzeeshangul.math@gmail.com}\\
Department of Mathematics, University of the Punjab,\\
Quaid-e-Azam Campus, Lahore-54590, Pakistan.}

\date{}
\maketitle

\begin{abstract}
In this paper, we investigate the newly developed
$f(R,\mathbf{T}^2)$ theory ($R$ is the Ricci scalar
and $\mathbf{T}^2=T_{\alpha\beta}T^{\alpha\beta},~T
_{\alpha\beta}$ demonstrates the energy-momentum
tensor) to explore some viable cosmological models.
For this purpose, we use the Noether symmetry approach
in the context of flat Friedmann-Robertson-Walker (FRW)
universe. We solve the Noether equations of this
modified theory for two types of models and obtain the
symmetry generators as well as corresponding conserved
quantities. We also evaluate exact solutions and
investigate their physical behavior via different
cosmological parameters.
For the prospective models, the graphical behavior of
these parameters indicate consistency with recent
observations representing accelerated expansion of the
universe. In the first case, we take a special model of
this theory and obtain new class of exact solutions with
the help of conserved quantities. Secondly, we consider
minimal and non-minimal coupling models of $f(R,\mathbf{T}
^{2})$ gravity. We conclude that conserved quantities are
very useful to derive the exact solutions that are used
to study the cosmic accelerated expansion.
\end{abstract}

\section{Introduction}

The accelerated expansion of the universe has been the most
unexpected and surprising result for the scientific community for
the last two decades. Since gravity being an attractive force will
lead the universe and all the matter present inside it to contract,
hence the expansion of the universe would gradually slow down.
However, this is against the observational evidences and hence we
need to search for some new physics which is consistent with our
observations. The well-known approach is modifying the geometry of
spacetime, i.e., general theory of relativity (GR) at large
distances, specifically beyond our solar system to produce
accelerating cosmological solutions \cite{1a}. Modified theories can
be formulated by adding the functions of curvature invariants in the
geometric part of the Einstein-Hilbert action. The natural
modification is obtained by replacing an arbitrary function of the
Ricci scalar $R$ in the Einstein-Hilbert action, so called $f(R)$
theory of gravity. There have been a crucial literature \cite{1}
available to understand the viable characteristics of this gravity.

The $f(R)$ theory of gravity has further been generalized by
introducing some couplings between the geometrical quantities and
the matter sector. The non-minimally coupling between the curvature
invariant and matter lagrangian density $(\mathcal{L}_{m})$ has been
established in \cite{9} dubbed as $f(R,\mathcal{L}_{m})$ theory of
gravity. These curvature-matter couplings explain various cosmic
eras as well as the rotation curves of galaxies. Such interactions
also include non-conserved energy-momentum tensor indicating the
existence of an additional force. These theories play a significant
role to understand the expanding behavior of the universe and dark
matter/energy interactions \cite{10}. One such modifications gave
rise to $f(R,T)$ theories ($T$ represents the trace of
energy-momentum tensor) \cite{6a}. A more generic theory in which
matter is nonminimally coupled to geometry was proposed \cite{6d},
referred to as $f(R,T,R_{\alpha\beta}T^{\alpha\beta})$ gravity
($R_{\alpha\beta}$ is the Ricci tensor and $T_{\alpha\beta}$ is the
energy-momentum tensor). Sharif and Ikram \cite{6} formulated such a
coupling in $f(\mathcal{G})$ gravity known as $f(\mathcal{G},T)$
theory, here $\mathcal{G}$ defines the Gauss-Bonnet invariant.
Moraes and Santos \cite{7} established $f(R, T^{\phi})$ theory,
where $T^{\phi}$ demonstrates the trace of the energy-momentum of
the scalar field.

This generalization procedure for the $f(R,\mathcal{L}_{m})$ theory
can also modify the corresponding Lagrangian by including some
analytic function of $T_{\alpha\beta}T^{\alpha\beta}$. This choice
of the corresponding Lagrangian will lead to
$f(R,T_{\alpha\beta}T^{\alpha\beta})$ theory of gravity, also called
energy-momentum squared gravity. Katirci and Kavuk \cite{6b}
proposed such a theory for the first time in 2014, which allows the
existence of a term proportional to $T_{\alpha\beta}T^{\alpha\beta}$
in the action functional. Different researchers have carried out
further studies on this theory. There has been a recent literature
\cite{13a} that indicates various cosmological applications of this
modified theory.

Roshan and Shojai \cite{15} found that this theory has a bounce at
early times and avoids the existence of singularity. Further, they
argued that the ``repulsive" nature of the cosmological constant
plays a significant role at early times for resolving the
singularity only after matter-dominated era. Board and Barrow
\cite{11b} investigated the range of exact solutions for isotropic
spacetime, presence of singularities, cosmic accelerated expansion
as well as evolution with a particular model of this theory. Morares
and Sahoo \cite{12a} studied non-exotic matter wormholes while
Akarsu et al. \cite{12b} explored possible constraints from neutron
stars in this framework. Bahamonde et al. \cite{13} studied
different cosmological models to investigate the ambiguous cosmic
characteristics. Akarsu et al. \cite{20a} investigated the minimal
and non-minimal curvature-matter coupling models of
$f(R,\mathbf{T}^{2})$ theory and observed that these models describe
the current cosmic accelerated expansion. Nari and Roshan \cite{17}
studied physical viability and stability of compact stars in this
framework. Bahamonde et al. \cite{20} studied dynamical system
analysis of this theory and found that this theory can explain the
current evolution of the universe and the emergence of the
accelerated expansion as a geometrical consequence. This literature
clearly motivates that $f(R,\mathbf{T}^2)$ gravity requires more
focus and there are many open issues that can be studied. This would
add and improve our current knowledge about different modified
theories of gravity.

Symmetry is a well-known significant aspect of cosmology as well as
theoretical physics. In this regard, Noether symmetry technique
helps to find exact solutions of the defined Lagrangian. It is an
interesting approach that suggests a correlation between conserved
quantities as well as symmetry generators of a dynamical system
\cite{21}. Such symmetries enable us to find analytical solutions of
nonlinear partial differential equations (PDEs) by reducing them to
a linear one. The main motivation comes from various conservation
laws (energy, momentum, angular momentum, etc.) which are outcomes
of some kind of symmetry being present in a system. The conservation
laws are the key factors in the study of various physical processes
and Noether theorem implies that every differentiable symmetry of
the action leads to the law of conservation. This theorem is
significant because it provides a correlation between conserved
quantities and symmetries of a physical system \cite{22}.
Capozziello and Ritis \cite{25} investigated the Noether symmetries
and also found the exact cosmological solutions in non-minimally
coupled gravitational theory. Capozziello et al. \cite{24} examined
Noether symmetry approach in the phantom quintessence universe.
Sharif and his collaborators \cite{38} analyzed the current cosmic
expansion and evolution by using this approach.

Capozziello et al. \cite{27} used Noether symmetry approach to find
static and non-static spherical solutions in $f(R)$ theory. Roshan
and Shojai \cite{34} studied Palatini $f(R)$ cosmology using Noether
symmetry approach for the matter-dominated universe. Hussain et al.
\cite{23} used this technique to analyze the Noether gauge symmetry
in the background of $f(R)$ theory. Shamir et al. \cite{28} applied
this symmetry approach to analyze the stability criteria of $f(R)$
gravity models for spherically symmetric as well as FRW universe.
Kucukakca et al. \cite{31} applied the Noether symmetry technique to
obtain analytic solutions of the Bianchi type-I spacetime. Shamir
and Ahmad \cite{37} discussed some cosmological models with
isotropic as well as anisotropic matter distribution though this
technique in $f(\mathcal{G},T)$ theory. Bahamonde et al. \cite{35}
used this approach to obtain various exact solutions of teleparallel
gravity with boundary term.

In this paper, we study the existence of Noether symmetry of flat FRW
universe in $f(R,\mathbf{T}^{2})$ theory of gravity. We determine
possible symmetries as well as corresponding conserved quantities
and evaluate exact solutions for two $f(R,\mathbf{T}^{2})$ models
to analyze cosmic evolution through cosmological parameters.
The paper is planned as follows. In section
\textbf{2}, we study some basic facts of this theory. Section
\textbf{3} gives a brief description about symmetry minimized
Lagrangian and Noether equations. Section \textbf{4} provides
cosmological solutions based on the conserved quantities. A brief
summary and discussion of the results is given in the last section.

\section{Basics of $f(R,\mathbf{T}^2)$ Gravity}

In this section, we formulate the field equations for
$f(R,\mathbf{T}^2)$ theory in the presence of perfect fluid. The
action for this gravity can be expressed as \cite{20}
\begin{equation}\label{1}
S=\frac{1}{2\kappa^2}\int f\left
(R,\mathbf{T}^{2}\right){\sqrt{-g}}d^4x+\int
\mathcal{L}_{m}\sqrt{-g}d^4x,
\end{equation}
where $\kappa^{2}$, $g$ and $\mathcal{L}_{m}$ represent the coupling
constant, determinant of the metric tensor and the Lagrangian
density of matter, respectively. We consider coupling constant as a
unity for the sake of simplicity. The action indicates that this
theory has extra degrees of freedom. Therefore, the possibility of
exact solutions is enhanced as compared to GR. Due to matter
dominated era, it is expected that some useful consequences would be
obtained to study the issues of dark energy and current cosmic
expansion in this gravity. The variation of the action with respect
to the metric tensor yields the following field equations
\begin{equation}\label{2}
R_{\alpha\beta}f_{R}+g_{\alpha\beta}\Box f_{R}-\nabla_{\alpha}
\nabla_{\beta}f_{R}-\frac{1}{2}g_{\alpha\beta}f =
T_{\alpha\beta}-\Theta_{\alpha\beta}f_{\mathbf{T}^{2}},
\end{equation}
where $\Box= \nabla_{\alpha}\nabla^{\alpha}$, $f\equiv f(R,
\mathbf{T}^{2})$, $f_{\mathbf{T}^{2}}= \frac{\partial f}{\partial
\mathbf{T}^{2}}$, $f_{R}= \frac{\partial f}{\partial R}$, and
\begin{eqnarray}\label{3}
\Theta_{\alpha\beta}
=-2\mathcal{L}_{m}\left(T_{\alpha\beta}-\frac{1}{2} g_{\alpha\beta}
T\right)-4\frac{\partial^{2}\mathcal{L}_{m}}{\partial
g^{\alpha\beta}\partial g^{\mu\nu}}T^{\mu\nu}
-TT_{\alpha\beta}+2T_{\alpha}^{\mu}T_{\beta\mu}.
\end{eqnarray}
For $f(R,\mathbf{T}^{2})= f(R)$, the field equations of this gravity
reduces to $f(R)$ theory and GR is recovered when
$f(R,\mathbf{T}^{2})=R$ \cite{40,41}.

We consider the matter configuration as a perfect fluid
\begin{equation}\label{4}
T^{m}_{\alpha\beta}=
(\rho+p)U_{\alpha}U_{\beta}+pg_{\alpha\beta},
\end{equation}
where $U_{\alpha}$, $p$ and $\rho$ depict the four velocity, energy
density and pressure, respectively. We assume the matter Lagrangian
as $\mathcal{L}_{m}=p$ so that
\begin{equation}\nonumber
\Theta_{\alpha\beta}=-U_{\alpha}U_{\beta}\left(4p\rho+\rho^{2}+3p^{2}\right).
\end{equation}
Rearranging Eq.(\ref{2}), we obtain
\begin{equation}\label{5}
G_{\alpha\beta}= R_{\alpha \beta }-\frac{1}{2}Rg_{\alpha \beta}=
\frac{1}{f_{R}}\left(T_{\alpha\beta}^{c}+T_{\alpha\beta}^{m}\right),
\end{equation}
where $T_{\alpha\beta}^{c}$ are the correction terms of
$f(R,\mathbf{T}^2)$ theory given as follows
\begin{equation}\label{6}
T_{\alpha\beta}^{c}=
\frac{1}{2}g_{\alpha\beta} \left(f-R
f_{R}\right)-g_{\alpha\beta}\Box
f_{R}+\nabla_{\alpha}\nabla_{\beta}f_{R}
-\Theta_{\alpha\beta}f_{\mathbf{T}^{2}}.
\end{equation}
Equation (\ref{5}) indicates that the stress-energy tensor of
gravitational fluid $(T_{\alpha\beta}^{c})$ gives matter contents
of the spacetime. Consequently, this technique includes all the
components of matter that might be significant to uncover the cosmic
mysteries. By contracting Eq.(\ref{2}), we have
\begin{equation}\label{7}
Rf_{R}-2f+3\Box f_{R}=T-\Theta f_{\mathbf{T}^{2}}.
\end{equation}
The flat FRW universe model is given by
\begin{equation}\label{8}
ds^{2}= -dt^{2}+a^2(t)\left(dx^{2}+dy^{2}+dz^{2}\right),
\end{equation}
where $a(t)$ defines the cosmic scale factor. The corresponding
dynamical quantities $R$ and $\mathbf{T}^{2}$ are
\begin{equation}\label{9}
R=6\frac{\dot{a}^{2}}{a^{2}}+6\frac{a\ddot{a}}{a^{2}},\quad
\mathbf{T}^{2}=3p^2+\rho^{2}.
\end{equation}
The respective field equations turn out to be
\begin{equation}\label{11}
-3\frac{\ddot{a}}{a}f_{R}+\frac{1}{2}f+3\frac{\dot{a}}{a}
\dot{f_{R}}= \rho-
\left(3p^{2}+\rho^{2}+4p\rho\right)f_{\mathbf{T}^{2}},
\end{equation}
\begin{equation}\label{12}
\left(\frac{\ddot{a}}{a}+2\frac{\dot{a}^{2}}{a^2}
\right)f_{R}-\frac{1}{2}f-\ddot{f_{R}}-2\frac{\dot{a}}{a}\dot{f_{R}}=p,
\end{equation}
where dot defines the rate of change with respect to time.

The field Eqs.(\ref{11}) and (\ref{12}) are highly non-linear as
well as complicated due to the presence of multivariate function and
its derivatives. In order to solve these equations, we consider the
Noether symmetry approach and determine exact solutions of
$f(R,\mathbf{T}^2)$ field equations. Since the conservation law does
not hold in this theory but we obtain conserved quantities in the
background of Noether symmetry approach. These are helpful to obtain
physically viable analytic or numeric solutions as well as to
analyze the mysterious universe. We analyze some feasible models of
cosmology through Noether symmetry approach.

\section{Symmetry Reduced Lagrangian and Noether Equations}

Noether symmetry provides a fascinating procedure to develop new
cosmological models and related geometries in modified gravitational
theories. Here, we formulate the point-like Lagrangian for FRW
universe in the background of $ f(R,\mathbf{T}^2)$ theory. We
determine the corresponding equations by using Noether symmetry
technique. This method provides a unique nature of the vector field
within the tangent space associated with it. Hence, the vector field
behaves as a symmetry generator and gives conserved quantities which
are then useful to examine exact solutions of the modified field
equations.

The canonical form of the action (\ref{1}) gives
\begin{equation}\label{13}
S= \int L\left(a,\dot{a},R,\dot{R},\mathbf{T}^{2},
\dot{\mathbf{T}^{2}}\right)dt.
\end{equation}
Using Lagrange multiplier approach, we have
\begin{equation}\label{14}
S= \int \sqrt{-g}\Big\{f-(R-\bar{R})\nu_{1}-
(\mathbf{T}^{2}-\mathbf{\bar{T}}^{2})\nu_{2}+p(a)\Big\}dt,
\end{equation}
where $\bar{R}=6\left(\frac{\dot{a}^{2}+a\ddot{a}}{a^{2}}\right)$,
$\mathbf{\bar{T}}^{2}=3p^2+\rho^{2}$ and $\sqrt{-g}= a^{3}$. We see that if
$R-\bar{R}=0$ and $\mathbf{T}^{2}-\mathbf{\bar{T}}^{2}=0$, then the
above action reduces to the action (\ref{1}) for FRW universe.
Varying Lagrange multipliers $\nu_{1}$ and $\nu_{2}$ with respect to
$R$ and $\mathbf{T}^{2}$, we obtain
\begin{equation}\label{15}
\nu_{1}= f_{R}, \quad \nu_{2}= f_{\mathbf{T}^{2}}.
\end{equation}
The corresponding action (\ref{14}) yields
\begin{equation}\label{16}
S=\int a^{3}\Big\{f-\left(R-6\frac{\dot{a}^{2}}{a^{2}}
-6\frac{a\ddot{a}}{a^{2}}\right)f_{R}-\left(\mathbf{T}^{2}-3p^2-\rho^{2}\right)f_{\mathbf{T}^{2}}+p\Big\}dt.
\end{equation}
Eliminating the boundary terms with the help of integration by
parts, we have
\begin{eqnarray}\nonumber
L(a,\dot{a},R,\dot{R},\mathbf{T}^{2},\dot{\mathbf{T}^{2}})&=&
a^{3}\left(f-R f_{R}-\mathbf{T}^{2} f_{\mathbf{T}^{2}}+
\left(3p^{2}+\rho^{2}\right)f_{\mathbf{T}^{2}}+p\right)
\\\label{17}&-&6a\dot{a}^{2}f_{R}-6a^{2} \dot{a}\left(\dot{R}
f_{RR}+\dot{\mathbf{T}^{2}}f_{R\mathbf{T}^{2}}\right).
\end{eqnarray}
The Euler-Lagrange equations is given by
\begin{equation}\label{18}
\frac{\partial L}{\partial q^{i}}-\frac{d}{dt}\left(\frac{\partial
L}{\partial \dot{q}^{i}}\right)=0, \quad i=1,2,3,...,n,
\end{equation}
where $q^{i}$ represent the generalized coordinates of
$n$-dimensional configuration space. By using
Lagrangian (\ref{17}), Eqs.(\ref{18}) turn out to be
\begin{eqnarray}\nonumber
&&\frac{1}{2}\left(f-R f_{R}-\mathbf{T}^{2} f_{\mathbf{T}^{2}}
+\left(3p^{2}+\rho^{2}\right)f_{\mathbf{T}^{2}}+p\right)
+2\frac{\dot{a}}{a}\dot{f_{R}}\\\label{19}&+&\frac{a}{6}\Big\{
\left(6pp_{,a}+2\rho\rho_{,a}\right)f_{\mathbf{T}^{2}}+p_{,a}\Big\}
+\left(\frac{\dot{a}^{2}}{a^{2}}+2\frac{\ddot{a}}{a}\right)f_{R}
+\ddot{f_{R}}=0,
\\\label{20}&&
\left(R-6\frac{\dot{a}^{2}}{a^{2}}-6\frac{\ddot{a}}
{a}\right)f_{RR}+\left(\mathbf{T}^{2}-3p^2-\rho^{2}\right)
f_{R\mathbf{T}^{2}}=0,
\\\label{21}&&
\left(R-6\frac{\dot{a}^{2}}{a^{2}}-6\frac{\ddot{a}}{a}
\right)f_{R\mathbf{T}^{2}}+\left(\mathbf{T}^{2}-3p^2-\rho^{2}\right)
f_{\mathbf{T}^{2} \mathbf{T}^{2}}=0.
\end{eqnarray}

The Hamiltonian of the Lagrangian is expressed as
\begin{equation}\label{22}
H= \dot{q}^{i}\left(\frac{\partial L}{\partial \dot{q}^{i}}
\right)-L.
\end{equation}
Using Eq.(\ref{17}), it turns out to be
\begin{eqnarray}\nonumber
H&=&-a^{3}\left(f-R f_{R}-\mathbf{T}^{2} f_{\mathbf{T}^{2}}
+\left(3p^{2}+\rho^{2}\right)f_{\mathbf{T}^{2}}+p\right)
\\\label{23}&-&
6a^{2}\dot{a}\left(\dot{R}f_{RR}+\dot{\mathbf{T}^{2}}f_{R\mathbf{T}^{2}}
\right)-6a\dot{a}^{2}f_{R}.
\end{eqnarray}
The generators of Lagrangian (\ref{17}) are considered as
\begin{equation}\label{24}
K= \tau\frac{\partial}{\partial t}+\xi^{i}\frac{\partial}{\partial
q^{i}},
\end{equation}
where $\tau\equiv \tau(t,a,R,\mathbf{T}^{2})$ and $\xi^{i}\equiv
\xi^{i}(t,a,R,\mathbf{T}^{2})$ for $i= 1,2,3,4$ are unknown
coefficients of the vector field $K$. The Lagrangian must fulfill
the condition of invariance for unique vector field $K$ over the
tangent space to assure the existence of Noether symmetries. In this
regard, the vector field acts as a symmetry generator that
constructs the conserved quantities. The invariance condition can be
expressed as
\begin{equation}\label{25}
K^{[1]}L+(D\tau )L= D\psi,
\end{equation}
where $\psi$ represents the boundary term, $K^{[1]}$ is the first
order prolongation and $D$ demonstrates the total derivative.
Further, it can be expressed as
\begin{equation}\label{26}
K^{[1]}= K+\dot{\xi}^{i}\frac{\partial}{\partial \dot{q}^{i}}, ~~~D=
\frac{\partial}{\partial t}+\dot{q}^{i}\frac{\partial}{\partial
\dot{q}^{i}},
\end{equation}
here $\dot{\xi}^{i}= D\xi^{i}-\dot{q}^{i}D\psi$.

The first integral of motion corresponds to Noether symmetry
generator $K$ determined as
\begin{equation}\label{27}
I= -\tau H+ \xi^{i}\frac{\partial L}{\partial\dot{q}^{i}}-\psi.
\end{equation}
This is the most significant part of Noether symmetries which is
also known as a conserved quantity. It is interesting to mention
here that the first integral plays a remarkable role to obtain
physically viable solutions. By considering Eq.(\ref{25}) and
comparing the coefficients, we obtain a set of PDEs as follows
\begin{eqnarray}\label{28}
&&\tau_{,a} =0,\quad \tau_{,R} =0,\quad \tau_{,\mathbf{T}^{2}}
=0, \quad \xi^{1}_{,R}f_{RR}=0,
\\\label{29}&&
6a^{2}\xi^{1}_{,t}f_{RR}+\psi_{,R}=0,\quad
6a^{2}\xi^{1}_{,t}f_{R\mathbf{T}^{2}}+\psi_{,\mathbf{T}^{2}}=0,
\\\label{30}&&
3\xi^{1}f_{\mathbf{T}^{2}}+a(\xi^{2}f_{R\mathbf{T}^{2}}+
\xi^{3}f_{\mathbf{T}^{2}\mathbf{T}^{2}})+a\tau_{,t}f_{\mathbf{T}^{2}}=0,
\\\label{31}&&
12a\xi^{1}_{,t}f_{R}+6a^{2}\left(\xi^{2}_{,t}f_{RR}+
\xi^{3}_{,t}f_{R\mathbf{T}^{2}}\right)+\psi_{,a}=0,
\\\nonumber&&
\xi^{1}f_{R}+a\left(\xi^{2}f_{RR}+\xi^{3}f_{R\mathbf{T}^{2}}\right)
+2a\xi^{1}_{,a}f_{R}-a\tau_{,t}f_{R}
\\\label{32}&&
+a^{2}\left(\xi^{2}_{,a}f_{RR}+\xi^{3}_{,a}f_{R\mathbf{T}^{2}}\right)=0,
\\\nonumber&&
2\xi^{1}f_{RR}+a\left(\xi^{2}f_{RRR}+\xi^{3}f_{RR\mathbf{T}^{2}}\right)
+2\xi^{1}_{,R}f_{R}-a\tau_{,t}f_{RR}
\\\label{33}&&
+a\left(\xi^{1}_{,a}f_{RR}+
\xi^{2}_{,R}f_{RR}+\xi^{3}_{,R}f_{R\mathbf{T}^{2}}\right)=0,
\\\nonumber&&
2\xi^{1}f_{R\mathbf{T}^{2}}+a\left(\xi^{2}f_{RR\mathbf{T}^{2}}
+\xi^{3}f_{R\mathbf{T}^{2}\mathbf{T}^{2}}\right)
+2\xi^{1}_{,\mathbf{T}^{2}}f_{R}-a\tau_{,t}f_{R\mathbf{T}^{2}}
\\\label{34}&&
+a\left(\xi^{1}_{,a}f_{R\mathbf{T}^{2}}+
\xi^{2}_{,\mathbf{T}^{2}}f_{RR}+\xi^{3}_{,
\mathbf{T}^{2}}f_{R\mathbf{T}^{2}}\right)=0,
\\\nonumber&&
3a^{2}\xi^{1}\Big\{f-R f_{R}-\mathbf{T}^{2}f_{\mathbf{T}^{2}}
+(3p^{2}+\rho^{2})f_{\mathbf{T}^{2}}+p\Big\}
\\\nonumber&&
+a^{3}\xi^{1}\Big\{(6pp_{a}+2\rho\rho_{a})f_{\mathbf{T}^{2}}
+p_{a}\Big\} + a^{3}\xi^{2}\Big\{-R f_{RR}-\mathbf{T}^{2}
f_{R\mathbf{T}^{2}}+
\\\nonumber&&
(3p^{2}+\rho^{2})f_{R\mathbf{T}^{2}}\Big\}+a^{3}\xi^{3}\Big\{-R
f_{R\mathbf{T}^{2}}-\mathbf{T}^{2}f_{\mathbf{T}^{2}\mathbf{T}^{2}}
+(3p^{2}+\rho^{2})f_{\mathbf{T}^{2}\mathbf{T}^{2}}\Big\}
\\\label{35}&&
+a^{3}\tau_{,t}\Big\{f-R f_{R}-\mathbf{T}^{2} f_{\mathbf{T}^{2}}
+(3p^{2}+\rho^{2})f_{\mathbf{T}^{2}}+p\Big\}-\psi_{,t}=0.
\end{eqnarray}
In the next section, we solve the above system of equations for
various cases.

\section{Conserved Quantities}

In this section, we manipulate the system of PDEs
(\ref{28})-(\ref{35}) to obtain Noether symmetries
$K=\tau\partial_{t} +\xi^{i}\partial_{q^{i}}$. Equation (\ref{28})
provides a trivial symmetry $I=\partial_{t}$ for any
$f(R,\mathbf{T}^{2})$ model. However, it is complicated to derive a
non-trivial solution without taking any particular
$f(R,\mathbf{T}^{2})$ model. In the following, we take different
models to reduce complexity of the system.

\subsection{$f(R)$ Gravity}

This case helps us to re-examine the usual $f(R)$ theory. The last
expression in Eq.(\ref{28}) implies that either $\xi^{1}_{,R} = 0$
or $f_{RR} = 0$. If we consider $f_{RR} = 0$ and $\xi^{1}_{,R} \neq
0$, then Eq.(\ref{33}) yields $f_{R} = 0$. Hence, Eq.(\ref{35}) with
$\tau = \tau(t)$ and $\psi= \psi(t)$ gives
\begin{equation}\label{36}
3a^{2}\xi^{1}(f+p)+a^{3}\tau_{,t}(f+p)+a^{3}\xi^{1}p_{a}-\psi_{,t}=0.
\end{equation}
Differentiating this with respect to $R$, we have $\xi^{1}_{,R} =
0$, that yields contradiction to the fact that $\xi^{1}_{,R} \neq
0$. So, our supposition is wrong and hence $f_{RR} \neq 0$ for
$f(R)$ theory. For the sake of convenience, we consider
$f(R,\mathbf{T}^{2})= f_{0}R^{\frac{3}{2}}$ \cite{39} which has
already been studied in the literature \cite{28},
\cite{40}-\cite{42}. By solving Eqs.(\ref{28}) to (\ref{35}), we
obtain
\begin{eqnarray}\label{37}
\xi^{1}=\frac{12c_{2}a^{2}f_{0}-2c_{1}c_{3}t-9c_{1}c_{5}f_{0}}
{18c_{1}a f_{0}}, \quad \xi^{2}= \left(\frac{2c_{3}t}{9a^{2}
f_{0}}-2\frac{c_{2}}{c_{1}}+\frac{c_{5}}{a^{2}}\right)R,
\\\label{38}
\tau= \frac{c_{2}t}{c_{1}}+c_{6}, \quad \psi= c_{2}t+c_{3}
a\sqrt{R}+c_{4}, \quad p= \frac{c_{1}}{a^{3}},
\quad \xi^{3}=\rho=0.
\end{eqnarray}
The Noether symmetry generators can be expressed as
\begin{eqnarray}\label{39}
K_{1}= -\frac{t}{9a f_{0}}\frac{\partial}{\partial a}+\frac{2tR}
{9a^{2}f_{0}}\frac{\partial}{\partial R}, \quad K_{2}=
-\frac{1}{2a}\frac{\partial}{\partial a}+\frac{R}{a^{2}}
\frac{\partial}{\partial R}, \quad K_{3}= \frac{\partial}{\partial
t}.
\end{eqnarray}
Using Eq.(\ref{27}), the conserved quantities can be found as
\begin{eqnarray}\label{40}
I_{1}&=& 9a\dot{a}^{2}\sqrt{R}f_{0}+\frac{9}{2}\frac{a^{2}\dot{a}
\dot{R}}{\sqrt{R}}f_{0}-\frac{a^{3}R^{3/2}}{2}f_{0},
\\\label{41}
I_{2}&=& \frac{a\dot{R}t}{2\sqrt{R}}+\dot{a}\sqrt{R}t-a\sqrt{R},
\quad I_{3}= \frac{9}{8}\frac{a\dot{R}t}{\sqrt{R}}+\frac{9}{2}
\dot{a}\sqrt{R}f_{0}.
\end{eqnarray}
These conserved quantities are the key aspects to determine the
cosmological solutions.

Now, we provide an important solution corresponding to the last
conserved quantity ($I_{3}$) that can be written as
\begin{equation}\label{42}
\dot{a}-\frac{2}{9\sqrt{R}f_{0}}\left(I_{3}-\frac{9a\dot{R}}
{4\sqrt{R}}f_{0}\right)=0.
\end{equation}
We can obtain a numerical solution by assuming $I_{3}= 2f_{0}$ with
some suitable initial conditions. The exact solution of
Eq.(\ref{42}) is of the following form
\begin{equation}\label{43}
a= R^{-1/2}\left(a_{0}+\frac{2I_{3}t}{9f_{0}}\right).
\end{equation}
Using the value of $R$ from Eq.(\ref{9}), the above equation
provides the exact solution for the cosmic scale factor
\begin{equation}\label{44}
a(t)=\sqrt{c_{0}+c_{1}t+c_{2}t^{2}+c_{3}t^{3}+c_{4}t^{4}},
\end{equation}
where $c_{i}$ are the combinations of initial conditions that help
to discuss cosmic evolution. If $c_{4}\neq 0$, then it gives a
power-law inflation while the radiation-dominated era is achieved
for the linear term in $c_{1}$ \cite{39}.

\subsection{$f(R,\mathbf{T}^{2})$ Gravity}

Here, we use curvature-matter coupling model to examine the
Noether symmetry technique in $f(R,\mathbf{T}^{2})$ theory. We
consider a specific type of a generic function both minimal as well
as non-minimal coupling between curvature and matter. We analyze the
cosmic evolution for the dust fluid.

\subsubsection{Minimal Coupling Models}

We take two minimal coupling models to find exact solutions. The
first minimal coupling model is given by \cite{17}
\begin{eqnarray}\label{44a}
f(R,\mathbf{T}^{2})= \alpha R^{n}+\beta(\mathbf{T}^{2})^{m}, \quad
n,m \neq 0,1,
\end{eqnarray}
where $\alpha$, $\beta$, $m$ and $n$ are constants. We consider
$\alpha=1=\beta$ for the sake of convenience. Solving the system
(\ref{28})-(\ref{34}), we have
\begin{eqnarray}\label{45}
\xi^{1}&=& ac_{2}+\frac{c_{1}}{a}, \quad \xi^{2}=
\frac{R^{2-n}c_{3}}{a}-\frac{R(3a^{2}c_{2}+c_{1})}{(n-1)a^{2}},
\\\label{46}
\xi^{3}&=& -\mathbf{T}^{2}\frac{3a^{2}c_{2}+3c_{1}}{(m-1)a^{2}},
\quad \tau= \psi =0.
\end{eqnarray}
Using Eqs.(\ref{45})-(\ref{46}) in (\ref{35}), we obtain
\begin{eqnarray}\nonumber
\rho &=&\Big\{R
c_{3}(\mathbf{T}^{2})^{1-m}\tan^{-1}\left(\frac{c_{2}a}{\sqrt{c_{1}c_{2}}}
\right)n\left(n-1\right)+R^{n}(\mathbf{T}^{2})^{1-m}
\\\nonumber&\times&
n\sqrt{c_{1}c_{2}}\left(2\ln(a)-\ln(c_{1}+a^{2}c_{2})\right)-3\mathbf{T}^{2}\ln(a)
\\\label{47}&\times&
\sqrt{c_{1}c_{2}}\left(R^{n}(\mathbf{T}^{2})^{-m}+1\right)\Big\}^{\frac{1}{2}}.
\end{eqnarray}
The symmetry generators take the following form
\begin{eqnarray}\label{48}
K_{1}&=& \frac{1}{a}\frac{\partial}{\partial a}-\frac{R}{(n-1)a^{2}}
\frac{\partial}{\partial R}-\frac{3\mathbf{T}^{2}}
{(m-1)a^{2}}\frac{\partial}{\partial \mathbf{T}^{2}},
\\\label{49}
K_{2}&=& a\frac{\partial}{\partial a}-\frac{3R}{(n-1)}\frac{\partial}
{\partial R}-\frac{3\mathbf{T}^{2}}{(m-1)}\frac{\partial}{\partial
\mathbf{T}^{2}}, \quad K_{3}=
\frac{R^{2-n}}{a}\frac{\partial}{\partial R}.
\end{eqnarray}
The corresponding first integrals become
\begin{eqnarray}\label{50}
I_{1}&=&
-6n\left(a\dot{R}R^{n-2}\left(n-1\right)+\dot{a}R^{n-1}\right), \quad
I_{2}= -6a\dot{a}n(n-1),
\\\label{51}
I_{3}&=&-6a^{2}n\left(a\dot{R}R^{n-2}(n-2)-\dot{a}R^{n}\right).
\end{eqnarray}

\begin{figure}
\epsfig{file=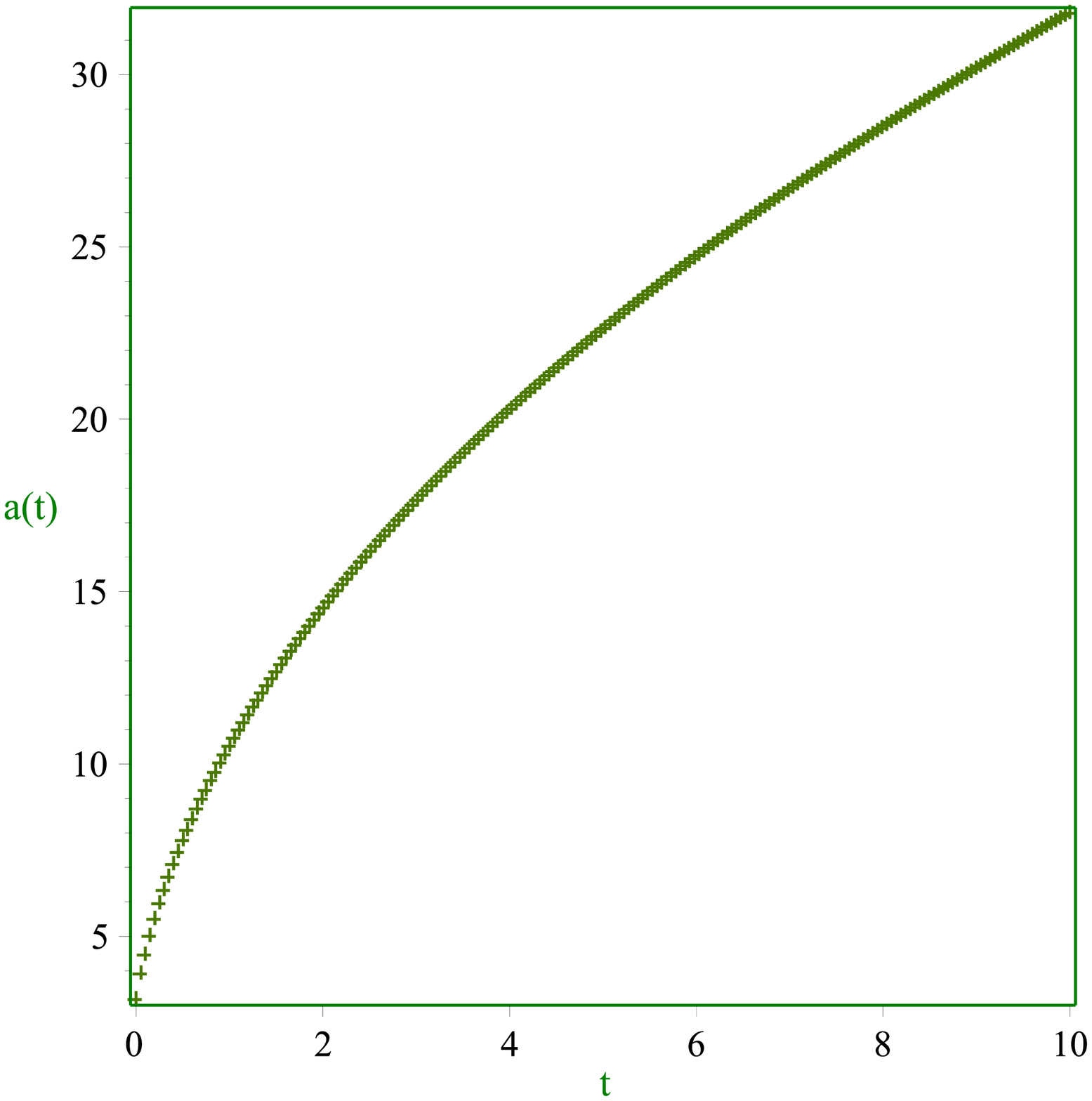,width=.5\linewidth}
\epsfig{file=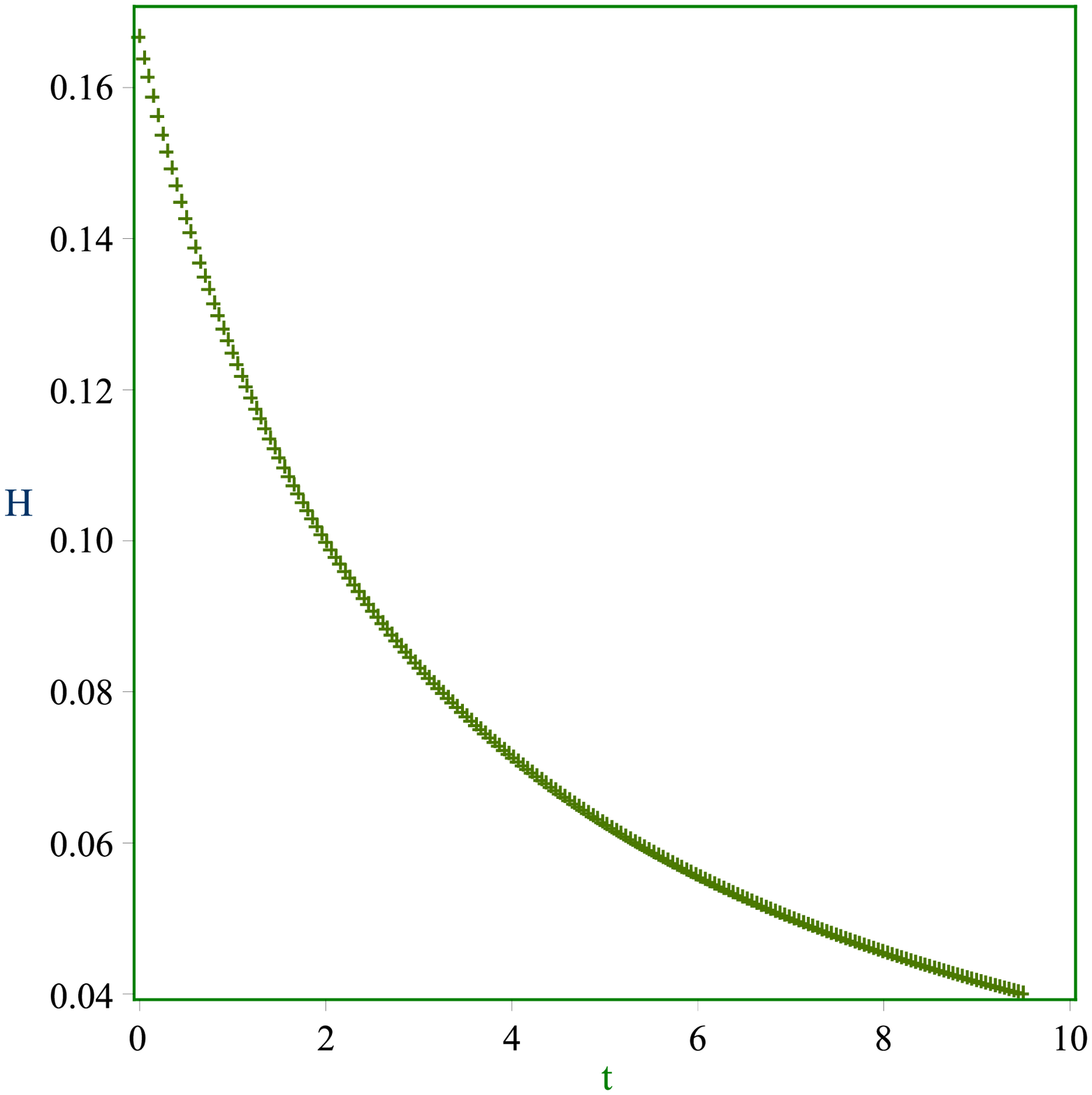,width=.5\linewidth} \caption{Plots of the scale
factor (left) and Hubble parameter (right) versus cosmic
time.}\label{52a}
\end{figure}

\begin{figure}\centering
\epsfig{file=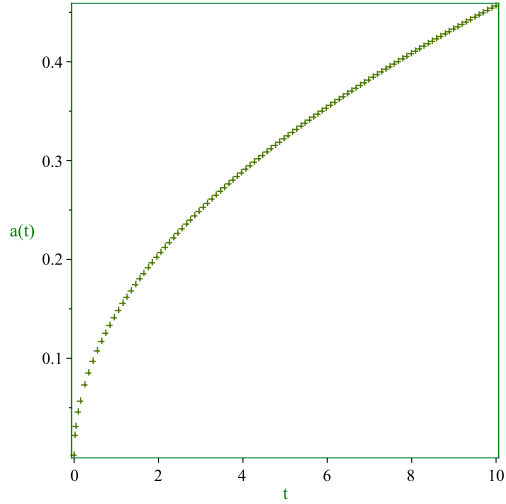,width=.5\linewidth} \caption{The evolutionary
behavior of the scale factor.}\label{52b}
\end{figure}

In order to establish cosmological analysis of the constructed model
experiencing minimal coupling with matter, we evaluate second
conserved quantity of Eq.(\ref{50}) as
\begin{equation}\label{50b}
a=\left(a_{0}-\frac{I_{2}t} {3n(n-1)}\right)^{\frac{1}{2}}.
\end{equation}
For the sake of simplicity, we consider $n=2$. Substituting the
value of second conserved quantity in the above equation, we obtain
exact solution of the scale factor as
\begin{equation}\label{50a}
a=\frac{\sqrt{c_{4}(c_{4}c_{5}-t)}}{c_{5}}.
\end{equation}
To investigate this solution, we discuss the behavior of some
significant cosmological parameters, i.e., Hubble, deceleration and
equation of state (EoS) parameters which play a crucial role in the
study of current accelerated expansion of the universe. The Hubble
parameter $(H)$ measures the rate of expansion whereas the value of
deceleration parameter $(q)$ determines accelerated $(q <0)$,
decelerated $(q>0)$, or constant expansion $(q = 0)$ of the
universe. For the isotropic universe model, the Hubble and
deceleration parameters are defined as
\begin{equation}\nonumber
H=\frac{\dot{a}}{a}, \quad q=-\frac{\dot{H}}{H^{2}}-1.
\end{equation}
For the explicit form of $f(R,\mathbf{T}^{2})$ model (\ref{44a}) and
scale factor (\ref{50a}), the corresponding Hubble and deceleration
parameters turn out to be $H=-\frac{1}{2(c_{1}c_{2}-t)}$ and $q=1$.
The graphical behavior of the scale factor and Hubble parameter is
shown in Figure \ref{52a}. The left plot indicates that the universe
experiences accelerated expansion as the scale factor grows
continuously while the right plot identifies decreasing rate of
expansion. The positivity of deceleration parameter ensures the
decelerating universe. Furthermore, the first integral (\ref{50})
provides a solution of the form
\begin{equation}\label{52}
a= R^{1-n}\left(a_{0}-\frac{I_{1}t}{6n}\right).
\end{equation}
The scale factor is physically viable due to its increasing
behavior, i.e., it describes cosmic accelerated expansion as shown
in Figure \ref{52b}. When $n=3/2$, this differential equation gives
the identical solution (\ref{43}) in $f(R)$ theory and GR is
recovered for $n=1$.

\begin{figure}
\epsfig{file=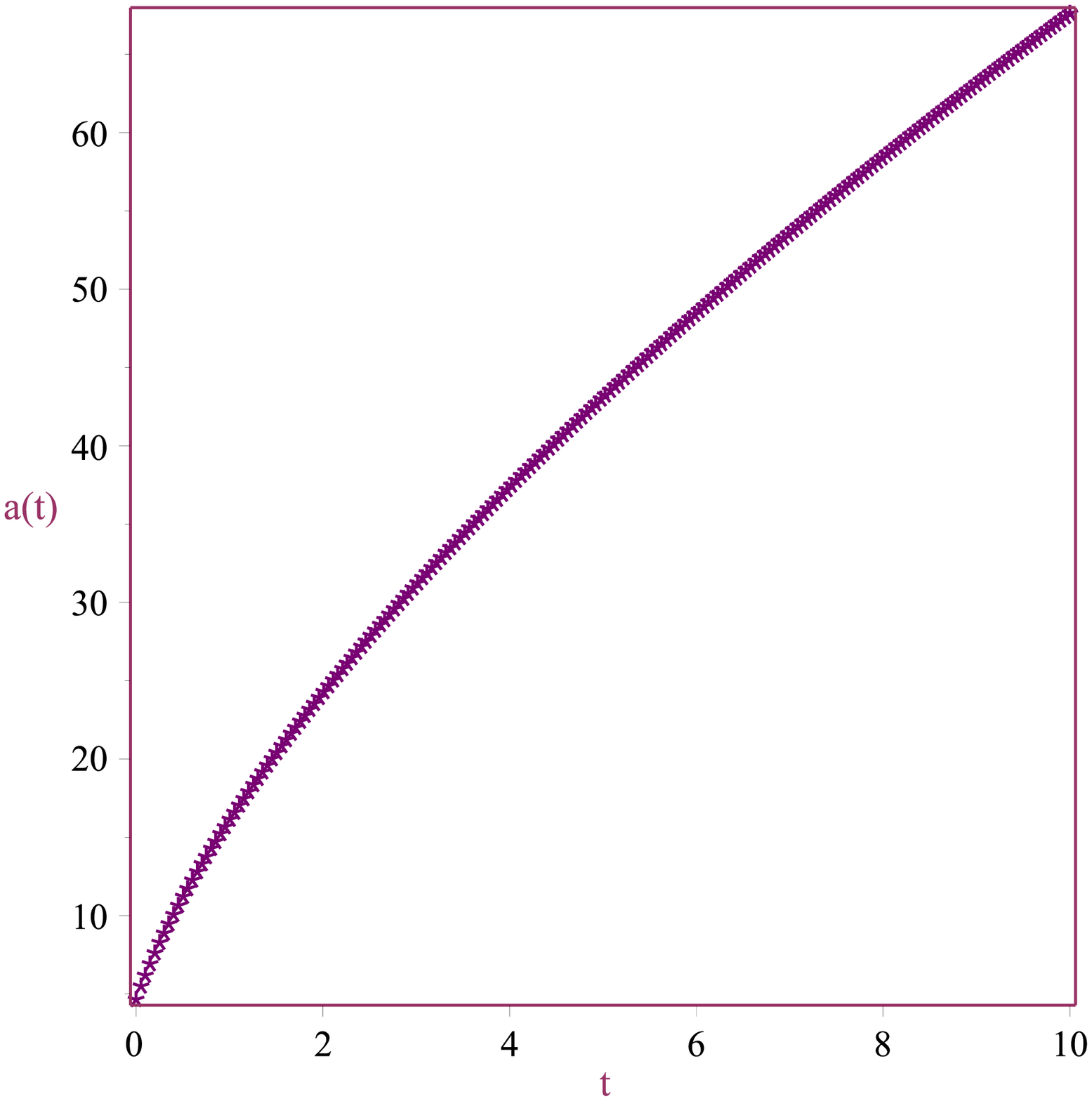,width=.5\linewidth}
\epsfig{file=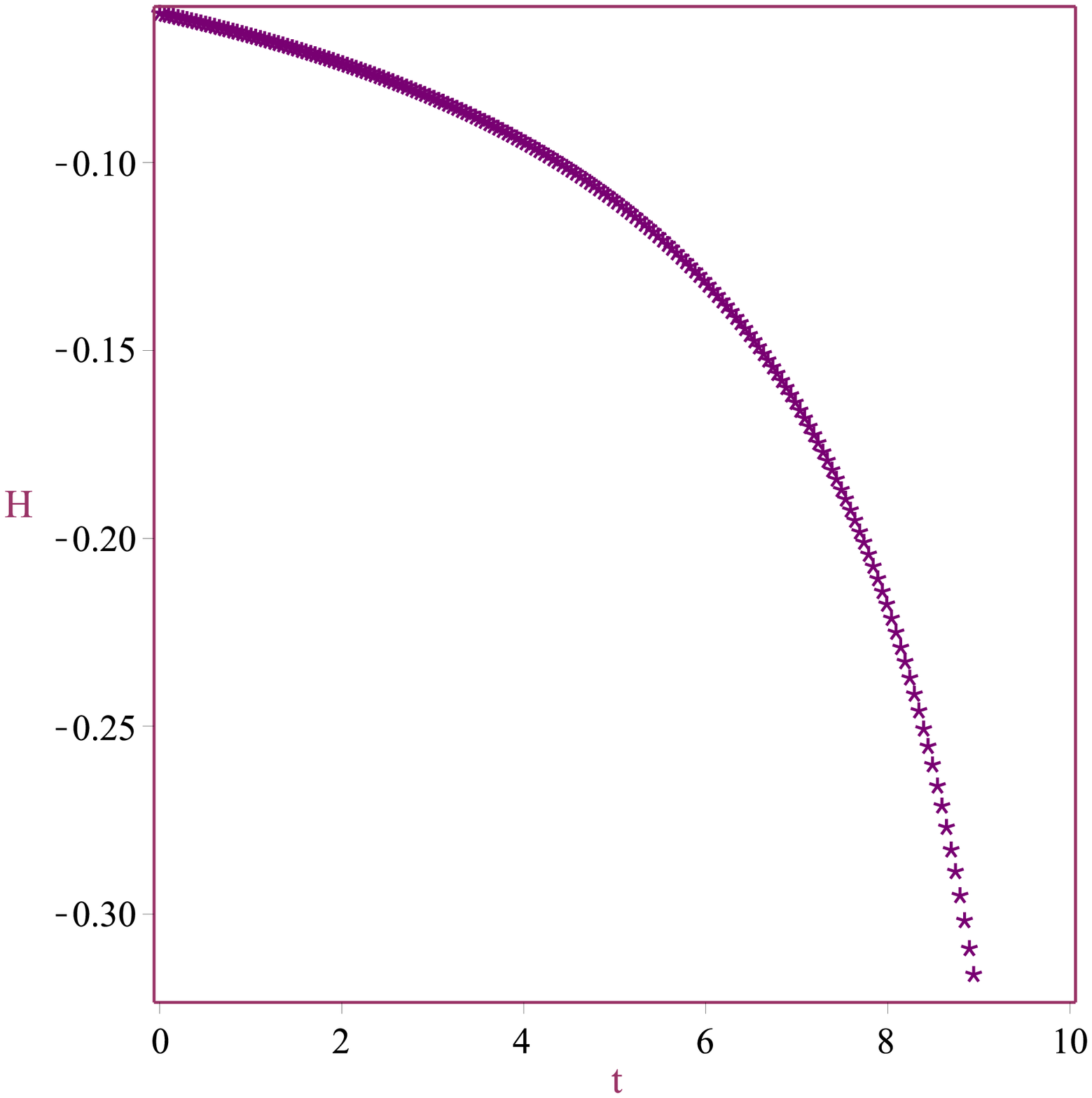,width=.5\linewidth} \caption{Plots of the scale
factor (left) and Hubble parameter (right) versus cosmic
time.}\label{52c}
\end{figure}

\begin{figure}\centering
\epsfig{file=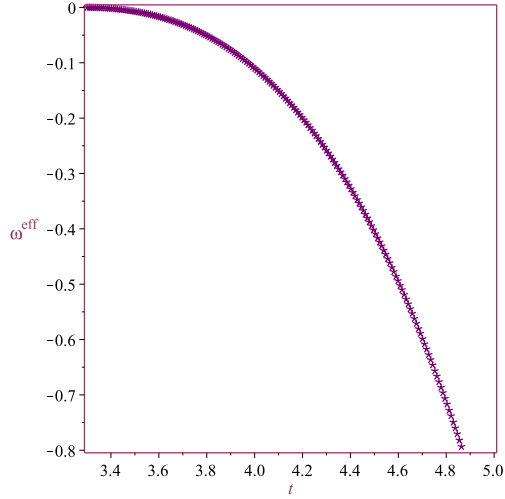,width=.5\linewidth} \caption{Plot of EoS
parameter versus cosmic time.}\label{52d}
\end{figure}

The second minimal coupling model is taken as
\begin{eqnarray}\nonumber
f(R,\mathbf{T}^{2})= R+\eta(\mathbf{T}^{2})^{n}, \quad n\neq 0,
\end{eqnarray}
where $\eta$ is a constant \cite{23}. The simultaneous solutions of
Eqs.(\ref{28})-(\ref{34}) yield
\begin{eqnarray}\label{53}
\xi^{1}= \frac{c_{1}}{\sqrt{a}}, \quad \xi^{3}=
-\frac{6\mathbf{T}^{2} c_{1}}{(n-1)a^{\frac{3}{2}}}, \quad
\xi^{2}=\tau=\psi=0.
\end{eqnarray}
Substituting these values in Eq.(\ref{35}), we have
\begin{equation}\label{54}
\rho= \sqrt{\frac{3\mathbf{T}^{2}\ln(a)}{n}}.
\end{equation}
The corresponding generators of the Noether symmetry and conserved
quantities become
\begin{eqnarray}\label{55}
K_{1}&=& \frac{1}{\sqrt{a}}\frac{\partial}{\partial a}-\frac{6
\mathbf{T}^{2}}{(n-1)a^{\frac{3}{2}}} \frac{\partial}{\partial
\mathbf{T}^{2}},
\\\label{56}
I_{1}&=&-12\dot{a}\sqrt{a},
\end{eqnarray}
respectively. Using Eq.(\ref{56}), we formulate exact solution of
the scale factor as
\begin{equation}\label{57a}
a=\frac{\{(c3e^{c2}+t)(e^{c2})^{2}\}^{\frac{2}{3}}}{(e^{c2})^{2}}.
\end{equation}
For this cosmological solution, Hubble and deceleration parameters
become $H=\frac{2}{3(c3e^{c2}+t)}$ and $q=\frac{1}{2}$,
respectively. The EoS parameter
$(\omega=\frac{p^{eff}}{\rho^{eff}})$ characterizes the universe
into different eras and also distinguishes DE era into distinct
phases like $\omega =-1$ describes cosmological constant, while
$-1<\omega \leq 1/3$ and $\omega < -1$ correspond to quintessence
and phantom phases, respectively. In Figure \ref{52c}, the left plot
indicates that the increasing behavior of the scale factor describes
accelerated expansion whereas the right plot represents that Hubble
parameter measures decreasing rate of cosmic expansion. The
positivity of deceleration parameter defines the decelerating
universe. Figure \ref{52d} shows that the universe possesses an
elegant exit from matter dominated era to phantom phase which leads
to quintessence phase with the passage of time.

\subsubsection{Non-Minimal Coupling Model}

This case explores the dynamical behavior with non-minimal coupling
model given as
\begin{eqnarray}\nonumber
f(R,\mathbf{T}^{2})= f_{0}R^{n}(\mathbf{T}^{2})^{m}, \quad n,m \neq
0,1,
\end{eqnarray}
where $f_{0}$ is a real constant \cite{17}. Solving
Eqs.(\ref{28})-(\ref{34}), we have
\begin{eqnarray}\label{58}
\xi^{1}&=& ac_{1}, \quad \xi^{2}=
\frac{1}{a}\Big\{\frac{-3Rac_{1}}{(n+m-1)}+R^{2-n}(
\mathbf{T}^{2})^{-m}c_{2}\Big\}, \quad \tau=0=\psi,\\\nonumber
\xi^{3}&=&
-\frac{\mathbf{T}^{2}}{\left(n+m-1\right)\left(m-1\right)a^{2}R}
\Big\{ac_{2}\left(n+m-1\right)nR^{2-n}(\mathbf{T}^{2})^{-m}
\\\label{59} &+&
3a^{2}(m-1)R c_{1}\Big\}.
\end{eqnarray}
Putting these values in Eq.(\ref{35}), we obtain
\begin{eqnarray}\nonumber
\rho&=&\Big\{-\frac{1}{m(m-1)a c_{1}}\{3\mathbf{T}^{2}a
c_{1}\left(m-1\right)\ln(a)
\\\label{60}&+&
(\mathbf{T}^{2})^{1-m}R^{1-n}nc_{2}\left(1-n-m\right)\}
\Big\}^{\frac{1}{2}}.
\end{eqnarray}
\begin{figure}\centering
\epsfig{file=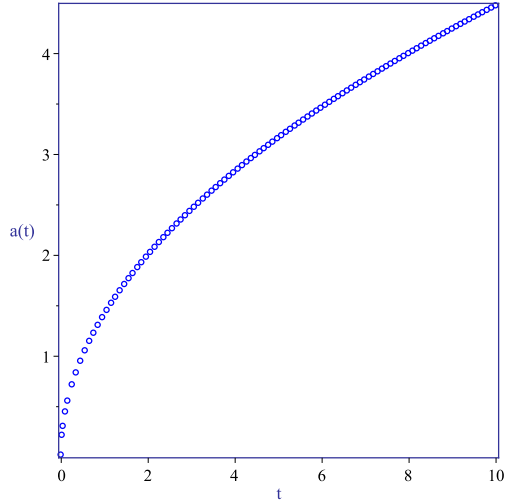,width=.5\linewidth} \caption{The evolutionary
behavior of the scale factor for
$f_{0}R^{n}(\mathbf{T}^{2})^{m}$.}\label{52e}
\end{figure}

The corresponding generators of Noether symmetry can be written as
\begin{eqnarray}\label{61}
K_{1}&=& a\frac{\partial}{\partial
a}-\frac{3R}{(n+m-1)}\frac{\partial} {\partial
R}-\frac{3\mathbf{T}^{2}}{(n+m-1)}\frac{\partial}{\partial \mathbf{T}^{2}},
\\\label{62}
\quad K_{2}&=& \frac{R^{2-n}}{(\mathbf{T}^{2})^{m}a}\frac{\partial}
{\partial R}+\frac{nR^{1-n}(\mathbf{T}^{2})
^{1-m}}{(1-m)}\frac{\partial}{\partial \mathbf{T}^{2}},
\end{eqnarray}
and the conserved quantities turn out to be
\begin{eqnarray}\nonumber
I_{1}&=& \frac{1}{(n+m-1)}\Big\{18a^{2}\dot{a}R^{n-1}
(\mathbf{T}^{2})^{m}(n^{2}-n+m-1)f_{0}\Big\}
\\\nonumber&+&
6a^{3}\dot{R}R^{n-2}(\mathbf{T}^{2})^{m}n(n-1)f_{0}-6a^{3}
nmR^{n-1}(\mathbf{T}^{2})^{m-1}\dot{\mathbf{T}^{2}}f_{0}
\\\label{63}&-&
12a^{2}\dot{a}nR^{n-1}(\mathbf{T}^{2})^{m}f_{0},
\\\label{64}
I_{2}&=&\frac{6\dot{a}mn^{2}}{(m-1)}f_{0}-6a\dot{a}n(n-1)f_{0}.
\end{eqnarray}
Equation (\ref{64}) can be rearranged as
\begin{equation}\label{65}
\dot{a}\Big\{\frac{6mn^{2}f_{0}}{(m-1)}-6an(n-1)f_{0}\Big\}-I_{2}=0.
\end{equation}
Using the initial condition $a(0)=0$ with $I_{2}=6f_{0}$, a
numerical solution is obtained. Figure \ref{52e} shows that the
scale factor describes the cosmic evolution for appropriate values
of $m$ and $n$. Equation (\ref{65}) gives analytic solution of the
form
\begin{equation}\label{66}
a=\Big\{a_{0}+\frac{I_{2}t(m-1)}{3n(n+m-1)f_{0}}\Big\}^{\frac{1}{2}}.
\end{equation}
When $m=0$, this equation provides the same solution as given for
the first minimal coupling model and reduces to GR if $m,~n=0$.
However, some interesting solutions can be established by taking
suitable values of $n$ and $m$.

\section{Concluding Remarks}

Modified gravitational theories are considered as the most
fascinating and promising approach to investigate the current cosmic
expansion due to the additional higher-order curvature terms. In
this paper, we have discussed Noether symmetries of
$f(R,\mathbf{T}^{2})$ theory for flat FRW universe model. Such
symmetries not only manage solutions of the dynamical system but
also their presence can provide some viable conditions so that
cosmological models can be selected according to current
observations \cite{43}. In particular, the characteristics of
mysterious energy associated with Noether symmetries can be
identified \cite{44}-\cite{47}. The Lagrangian multipliers are used
to minimize the dynamical system that ultimately help to evaluate
analytical solutions. We have formulated the Lagrangian of
$f(R,\mathbf{T}^{2})$ gravity and evaluated the conserved quantities
to investigate the exact solutions of modified equations of motion.
The analytic solutions of Noether equations have been studied for
minimal and non-minimal coupling models of this theory by assuming
dust fluid just for the sake of simplicity. We summarize the results
obtained as follows.
\begin{itemize}
\item Firstly, we have discussed exact solutions of Noether equations
for $f(R)$ model. The $f(R)$ theory is recovered for $f_{RR}\neq 0$.
We have considered $f(R,\mathbf{T}^{2})= f_{0}R^{3/2}$ with
$f_{0}\neq 0$ and derived Noether symmetries that are consistent
with those already present in the literature \cite{39,43}. We have
applied the conserved quantities to analyze numerical as well as
exact solutions for cosmic evolution. We have then formulated a
numeric solution after applying some suitable initial condition with
appropriate values of the parameters. The scale factor indicates
that the universe is expanding with an accelerating phase (Figure
\ref{52a}). The analytical approach provides an exact solution for
$f(R)$ gravity model \cite{40,45}.
\item There has been a significant literature \cite{48}-\cite{50} that
indicates various cosmological applications corresponding to this
cosmological model. Newtonian gravity is the weak-field limit of
general relativity and Modified Newtonian Dynamics (MOND) is the
weak-field limit of a particular extended theory of gravity. It has
been found that Noether symmetry approach yields a conserved
quantity coherent with the relativistic MONDian extension. The MOND
regime can be fully recovered as the weak-field limit of a
particular theory of gravity formulated in the metric approach. This
is possible when Milgrom's acceleration constant is taken as a
fundamental quantity which couples to the theory in a very
consistent manner. The power-law $f(R)$ gravity model demonstrates
the existence of a new fundamental gravitational radius. This radius
plays an analog role for weak gravitational field at galactic scales
and using the new radius, $f(R)$ gravity provides a theoretical
foundation for rotation curve of galaxies as well as empirical
baryonic Tully-Fisher relation. In particular, for $f(R)=R^{3/2}$,
the MOND acceleration regime is recovered.
\item In cosmology, perfect fluid can represent the effective behavior of
Hubble flow ranging from inflation to dark energy epochs. Therefore,
compatibility of perfect fluid solutions with modified or extended
theories of gravity is a crucial issue to be investigated. The
$n$-dimensional generalized Robertson-Walker spacetime with
divergence-free conformal curvature tensor exhibits a perfect fluid
stress-energy tensor for any $f(R)$ gravity model. Furthermore, a
conformally flat generalized Robertson-Walker spacetime is still a
perfect fluid in both $f(R)$ and quadratic gravity \cite{51}.
\item Secondly, we have studied minimal and non-minimal curvature-matter
coupling models of this theory. We have taken two minimal and one
non-minimal models. For the first minimal model,
$f(R,\mathbf{T}^{2})= \alpha R^{n}+\beta(\mathbf{T}^{2})^{m},~n,m
\neq 0,1$, we have obtained three conserved quantities out of which
two give a new framework of analytic solutions. In this case, we
have found cosmological solution of the scale factor whose physical
interpretation is established through cosmological parameters like
Hubble, deceleration and EoS parameters. The graphical analysis of
scale factor and rate of expansion is found to be increasing. The
deceleration parameter remains negative. The EoS parameter
characterize phantom phase which leads to quintessence phase with
the passage of time. For the second minimal model
$f(R,\mathbf{T}^{2}) = R+\eta(\mathbf{T}^{2})^{n}$ with $n\neq0$, we
have different solutions using conserved quantities for different
values of $n$. For the non-minimal model $f(R,\mathbf{T}^{2})=
f_{0}R^{n}(\mathbf{T}^{2})^{m}$, we have found two generators. It is
clear that the scale factor is rapidly increasing which indicates
the cosmic accelerated expansion for all cases (Figures
\ref{52a}-\ref{52e}).
\item It is worthwhile to mention here that the results of this theory are compatible with each other. In the
first minimal coupling model, the solution of scale factor
(\ref{52}) becomes $a=R^{-1/2}\left(a_{0}-\frac{I_{1}t}{9}\right)$
for $n=3/2$ which is similar to that obtained in $f(R)$ theory (42).
Similarly, GR is recovered when $n=1$. For non-minimal coupling
model, we have the solution of scale factor (\ref{66}) as
$a=\left(a_{0}+\frac{I_{2}t(m-1)}{3n(n+m-1)f_{0}}\right)^{\frac{1}{2}}$
which reduces to $a=\left(a_{0}-\frac{I_{2}t}
{3n(n-1)}\right)^{\frac{1}{2}}$ for $m=0$. It is identical to the
one discussed in (52). Substituting $m,~n=0$ in Eq.(\ref{66}) the
scale factor of GR is recovered.
\end{itemize}
We would like to mention here that our results reduce to some other
models of $f(R,\mathbf{T}^{2})$ gravity for different values of
parameters.

\end{document}